\documentclass[letterpaper]{article} 
\usepackage{aaai2026}  
\usepackage{times}  
\usepackage{helvet}  
\usepackage{courier}  
\usepackage[hyphens]{url}  
\usepackage{graphicx} 
\usepackage{natbib}  
\usepackage{caption} 
\usepackage{algorithm}
\usepackage{algorithmic}
\usepackage{xcolor}

\usepackage{booktabs}   
\usepackage{multirow}   
\usepackage{array}      
\usepackage{caption}    
\usepackage{longtable}  
\usepackage{tabularx}   
\usepackage{comment}    
\usepackage{amsmath}    
\usepackage{graphicx}   
\usepackage{enumitem}   
\usepackage{tcolorbox}  
\usepackage{natbib}     

\usepackage{xcolor}

\usepackage{inconsolata}

\usepackage{newfloat}
\usepackage{listings}
\DeclareCaptionStyle{ruled}{labelfont=normalfont,labelsep=colon,strut=off} 
\floatstyle{ruled}
\newfloat{listing}{tb}{lst}{}
\floatname{listing}{Listing}
\title{ParsCN: A Persian Dataset for Counter-Narrative Generation to Combat Online Hate Speech}
\author {
    Zahra Safdari Fesaghandis\textsuperscript{\rm 1},
    Suman Kalyan Maity\textsuperscript{\rm 2}
}
\affiliations {
    \textsuperscript{\rm 1}Bilkent University, Ankara, Turkey\\
    \textsuperscript{\rm 2}Missouri University of Science and Technology, Rolla, MO, USA\\
    \textsuperscript{\rm 1}zahra.safdari@bilkent.edu.tr, \textsuperscript{\rm 2}smaity@mst.edu
}

\begin{document}

\maketitle

\begin{abstract}
Online hate speech threatens online civility, particularly in low-resource and multilingual environments. Counter-narratives offer a promising solution by promoting constructive responses to hate speech. However, automatic counter-narrative generation is hindered by the lack of high-quality data for low-resource languages like Persian. To bridge this gap, we introduce ParsCN, the first and most comprehensive Persian counter-narrative dataset. Consisting of 1,100 hate speech and counter-narrative pairs, it provides fine-grained annotations across six target groups and six countering strategies, tailored to the socio-cultural context of Persian online discourse. We propose a novel, scalable multi-stage framework that integrates culturally-informed human annotation with few-shot LLM-augmented generation, guided by semantic retrieval and rigorous manual curation. This approach enables the creation of diverse, high-quality counter-narratives while significantly reducing annotation costs—establishing a replicable paradigm for other low-resource settings. Comprehensive human and automatic evaluations confirm the quality of the dataset and the effectiveness of the generated responses. Human-written counter-narratives achieved the highest scores for relevance (4.23), Effectiveness (4.21), fluency (4.92), and tone appropriateness (4.79), with GPT-4o and Claude closely following. Automatic evaluations show strong semantic alignment (BERTScore F1 up to 0.709), high lexical diversity, and low toxicity across all sources. Finally, we conduct benchmark evaluations using mBART and PersianMind on a held-out test set. Results reveal that existing models struggle with fluency, cultural nuance, and safety—highlighting the need for Persian-specific resources like ParsCN. Our dataset (available at \url{https://doi.org/10.5281/zenodo.18266930}) serves as a foundational benchmark to advance research on Persian counter-narrative generation and foster safer, more inclusive digital spaces.

\end{abstract}


\section{Introduction}
\label{sec:introduction}
Social media has profoundly reshaped human communication, facilitating global exchange while simultaneously magnifying challenges such as the spread of hate speech. Defined as language discriminating against individuals or groups based on ethnicity, religion, or gender, hate speech poses significant social and technical challenges, given the diversity of cultural norms, legal frameworks, and linguistic structures~\citep{warner2012detecting, watanabe2018hate, fortuna2018survey}. Traditional moderation methods like automated detection and removal face inherent limitations, including scale, censorship concerns, and difficulty capturing nuanced, evolving hateful language \citep{hee2024recent}.

To address these limitations, counter-narratives have emerged as a proactive and non-confrontational alternative. They involve non-aggressive responses that challenge hateful narratives and promote civil discourse \citep{chung2019conan, mathew2019thou}, recognized as a scalable means to cultivate inclusive digital spaces. However, a major barrier to automating counter-narrative generation—especially for languages beyond English—is the scarcity of high-quality, task-specific datasets. This gap is particularly critical for low-resource languages like Persian, where distinct linguistic and socio-cultural factors shape both hate speech and effective counter responses.

We present ParsCN, a novel dataset for generating counter-narratives to mitigate hate speech in Persian. 
Our key contributions are as follows:
\begin{itemize}
\item \textbf{ParsCN Dataset:} We present and publicly release ParsCN, the first dedicated dataset for counter-narrative generation in Persian. The dataset comprises 1,100 hate speech–counter-narrative pairs, annotated across six target groups and six distinct counter-narrative strategies, offering a nuanced and structured representation of online hate and its potential responses.
\item \textbf{Addressing the Low-Resource Gap:} It addresses the vast resource gap for counter-narrative generation in low-resource languages, providing a high-quality, foundational dataset that is particularly well adapted to the unique linguistic and socio-cultural context of Persian online discourse.  
\item \textbf{Novel Multi-Stage Framework:} We present and utilize a novel multi-stage framework for dataset creation that effectively leverages expert human annotation with the assistance of state-of-the-art large language models (LLMs) and automated methods, thereby proposing a scalable framework for the creation of comparable resources.
\end{itemize} 




By releasing ParsCN, we aim to catalyse further research in automated counter-narrative generation for low-resource settings, contributing to the broader effort of fostering safer and more inclusive digital environments.

\section{Related Work}

Counter-narrative generation is an active research area focused on combating online hate speech. This section reviews key datasets and generation methodologies, highlighting the novelty of our multi-stage framework for ParsCN in the context of existing approaches.

\subsection{Counter-Narrative Dataset Creation}

Creating high-quality datasets is fundamental. Early efforts like CONAN \citep{chung2019conan} pioneered multilingual (English, French, Italian) expert-based datasets using nichesourcing. MultiTarget-CONAN \citep{fanton2021human} introduced a human-in-the-loop approach for multi-target hate speech. More recently, datasets have expanded linguistically and more comprehensively, including IndicCONAN \citep{sahoo2024indicconan} for Indian languages, PANDA \citep{bennie2025panda} as the first Chinese dataset leveraging LLM-as-a-Judge, and CONAN-EUS \citep{bengoetxea2024basque} for Basque and Spanish via translation and post-editing, demonstrating models for low-resource contexts \citep{sahoo2024indicconan, bengoetxea2024basque}. Diverse sourcing methods like crowdsourcing \citep{saha2024crowdcounter} and utilizing in-the-wild data from platforms like YouTube \citep{fialho2024counter} have provided varied resources. Datasets focusing on specific aspects like counter-narrative types \citep{saha2024crowdcounter, sahoo2024indicconan} and intent \citep{hengle2024intent} have also been developed. Despite these advances, a significant gap remains for many languages, particularly low-resource datasets like Persian.

ParsCN’s novelty extends beyond being the first Persian CN dataset. It introduces: (1) a culturally tailored annotation framework addressing Persian-specific hate speech (e.g., targeting Afghan immigrants or Iran’s regime); (2) fine-grained annotations across six target groups and six CN strategies; (3) a hybrid data sourcing approach combining Persian PHATE data with translated MultiTarget-CONAN and HateXplain instances, rigorously post-edited for cultural relevance; and (4) a scalable multi-stage framework integrating culturally informed human annotation, semantic retrieval using a SentenceTransformer model for few-shot LLM prompting, and rigorous manual curation of LLM outputs. Unlike simpler pipelines in IndicCONAN (direct human annotation) or CONAN-EUS (translation-based), ParsCN’s framework prioritizes cultural relevance and scalability for low-resource settings. A detailed comparison with other low-resource CN datasets is provided in Appendix C, Table \ref{tab:dataset_comparison}.

\subsection{Counter-Narrative Generation Methodologies}
Various techniques have been explored for automated generation. Approaches leveraging external knowledge grounding \citep{chung2021towards, wilk2025fact} aim for more informative responses. Methods incorporating structural aspects like argumentative \citep{furman2023high} or discourse structure \citep{hassan2023discgen} seek better control and quality. Constraint-based generation, such as optimizing for desired conversation outcomes using LLMs \citep{mun2023beyond}, addresses real-world impact. LLMs are central to many systems \citep{podolak2023llm}, including retrieval-augmented approaches like ReZG \citep{jiang2025rezg} for zero-shot generation. Research also analyzes the use and effectiveness of specific strategies for countering implied biases \citep{mun2023beyond}. Furthermore, developing multilingual models and context-aware approaches \citep{bennie2025codeofconduct} is crucial for extending capabilities, especially in low-resource settings.

\section{Dataset}
\label{sec:Dataset}

\begin{figure*}[h!]
    \centering
\includegraphics[width=\textwidth]{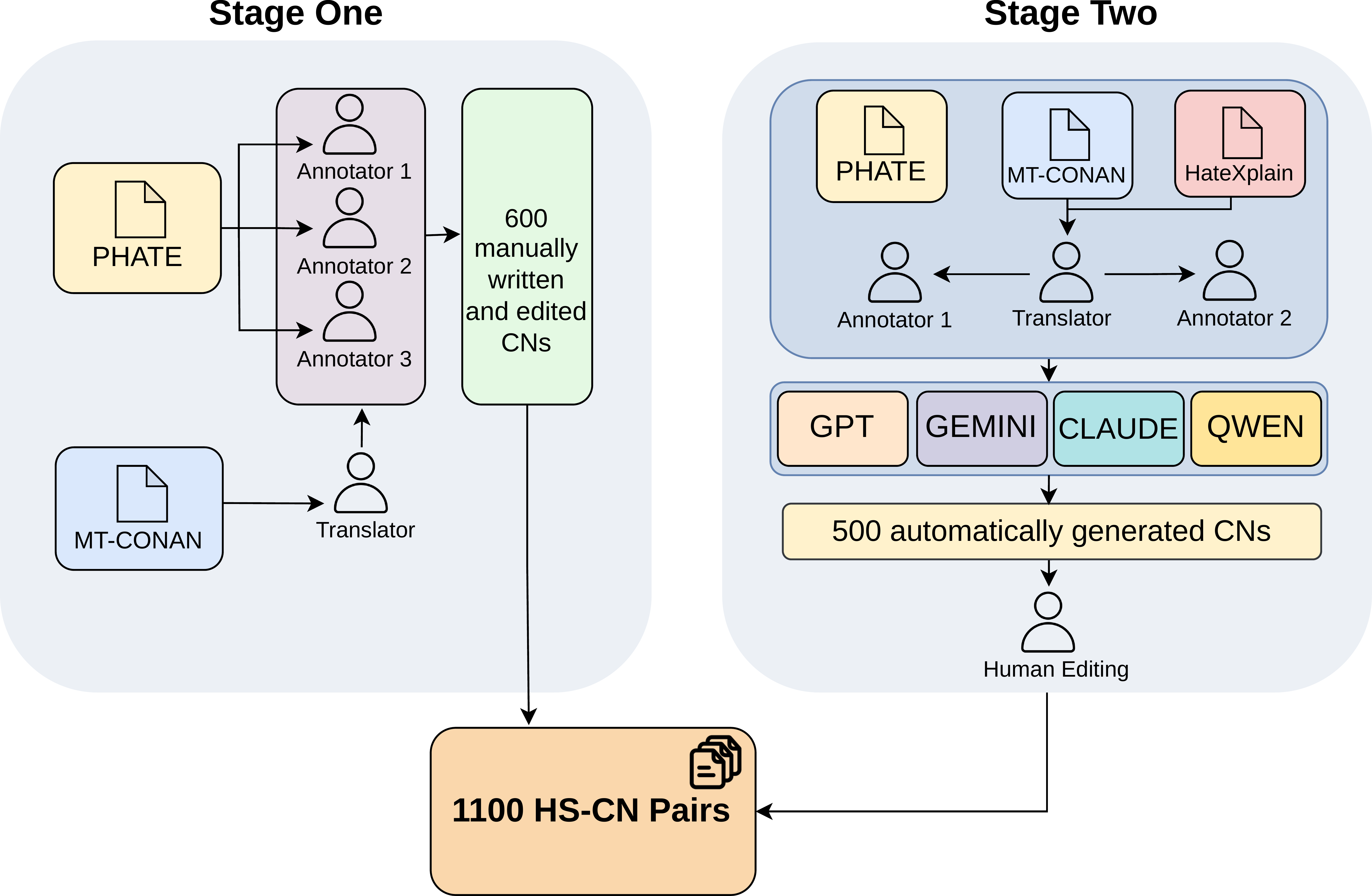}
    \caption{ParsCN Dataset Creation Pipeline}
    \label{fig:pipeline}
\end{figure*}

In this section, we present our \textbf{ParsCN} dataset designed to facilitate counter-narrative research in the Persian language, addressing the pressing issue of online hate speech in Persian-speaking communities. This dataset comprises pairs of hate speech and corresponding counter-narratives, annotated with hate speech target categories and counter-narrative types. \textbf{ParsCN} is the first dataset of its kind in Persian, filling a critical gap in resources for non-English counter-narrative research. Below, we describe the dataset’s characteristics, the two-stage process of data collection involving manual and automated annotation, and the annotation strategies employed to ensure quality and diversity. Figure~\ref{fig:pipeline} illustrates the two-stage process followed for the creation of the \textbf{ParsCN} dataset, combining data sourcing, human annotation, and automated generation techniques. 

\subsection{Hate Speech Datasets}
\label{subsec:hate_speech_datasets}
To construct \textbf{ParsCN}, we sourced hate speech instances from three publicly available datasets, adapting them to Persian linguistic and cultural contexts:

\textbf{PHATE:} A Persian dataset of 7,000 manually annotated tweets labeled as normal or hate (subcategorized as violence, hate, or vulgar) \citep{delbari2024spanning}. Annotations include target groups and justification spans, making it directly applicable to ParsCN.

\textbf{MultiTarget-CONAN:} An English dataset of $\sim$5,000 hate speech–counter-narrative pairs covering race, religion, and gender \citep{fanton2021human}. Created using a human-in-the-loop process with GPT-2 and NGO experts, it includes Persian-relevant targets (e.g., Muslims). We translated and manually validated culturally appropriate instances.

\textbf{HateXplain:} A 20,148-post English dataset from Twitter and Gab, annotated as hate, offensive, or normal \citep{mathew2021hatexplain}. We selected posts targeting Persian-relevant groups (e.g., ethnic and occupational), translating and validating them for cultural fit.

Together, these datasets provided a diverse pool of hate speech instances, which we filtered, translated, and annotated to construct ParsCN’s hate–counter-narrative pairs.

\subsection{Hate Speech Categories}
\label{subsec:hate_speech_categories}

\begin{figure*}[ht]
  \centering
  \includegraphics[width=0.95\textwidth]{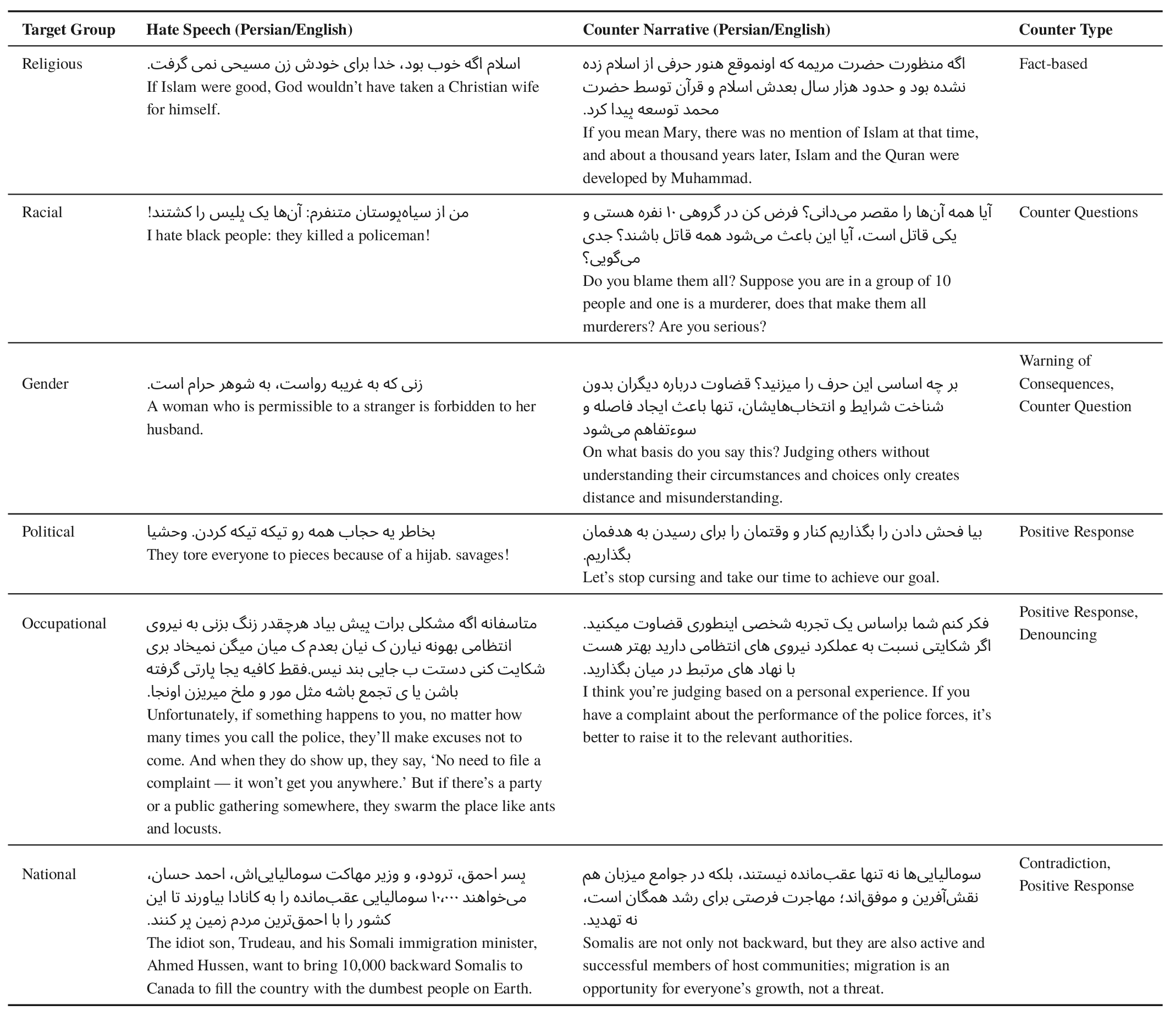}
  \caption{Examples of Hate Speech and Counter Narratives, Along with Their Target Groups and Counter Types, from ParsCN}
  \label{fig:examples}
\end{figure*}

To enable fine-grained analysis and generation of counter-narratives, we categorize hate speech in \textbf{ParsCN} into six types, directly adopted from the PHATE dataset \citet{delbari2024spanning} and tailored to the socio-cultural context of Persian-speaking regions~\citep{delbari2024spanning}. These categories are as follows - \textbf{i) Religious}: Statements targeting religious groups, such as Islam or the Jewish religion; \textbf{ii) Racial}: Speech discriminating against ethnic/racial groups like Azeris, Kurds, or Black Africans; \textbf{iii) Gender}: Statements derogatory to men, women, and including activists; \textbf{iv) Political}: Expressions targeting the government, politicians, or country laws; \textbf{v) Occupational}: Statements targeting professions, such as police or teachers; \textbf{vi) National}: Speech targeting national or immigrant groups like Afghans, Chinese, or Arabs.
Figure~\ref{fig:examples} provides one example per category, showcasing the dataset’s diversity. 

\subsection{Counter-Narrative Approaches}
\label{subsec:counter_narrative_approaches}
To produce effective counter-narratives against hate speech in Persian online platforms, \textbf{ParsCN} incorporates a diverse set of six counter-narrative strategies, adapted from \citep{sahoo2024indicconan} (see Table~\ref{tab:counter_narrative_types} and Appendix B). These strategies are tailored to address hate speech in six target categories: Religious, Racial, Gender, Political, Occupational, and National. Each type, represented in \textbf{ParsCN}’s dataset (see Figure~\ref{fig:examples} for a few examples from our dataset), ensures contextually appropriate counter-narratives for Persian hate speech.

\begin{table}[h]
\centering
\small
\caption{Summary of Counter-Narrative Types in ParsCN}
\label{tab:counter_narrative_types}
\resizebox{.95\columnwidth}{!}{
\begin{tabular}{>{\raggedright\arraybackslash}m{3cm}>{\raggedright\arraybackslash}m{4cm}}
\toprule
\textbf{Type} & \textbf{Purpose} \\
\midrule
Positive Response & Fosters empathy and inclusion \\
Counter questions & Prompts reflection on biases \\
Denouncing & Rejects harmful rhetoric \\
Fact-based & Corrects misinformation with evidence \\
Warning of consequences & Highlights negative outcomes \\
Contradiction & Exposes logical inconsistencies \\
\bottomrule

\end{tabular}}

\end{table}

\subsection{Stage One: Data Collection and Annotation}
\label{subsec:stage_one}

In the first stage of \textbf{ParsCN} development, we constructed a dataset of 600 hate speech--counter-narrative pairs to facilitate automated counter-narrative generation in Persian across the target categories. This phase involved curating a diverse set of hate speech instances, generating contextually appropriate counter-narratives based on the PHATE dataset, and incorporating translated examples from MT-CONAN. All entries were annotated according to the six counter-narrative strategies outlined in the Counter-Narrative Approaches section. The resulting dataset--balanced with 100 instances per category--provides a robust foundation for model training in Stage Two. A summary of the dataset composition is provided in Table~\ref{tab:dataset_composition}, with further methodological details described below.

\begin{table}[h]
\centering
\small
\caption{ParsCN Dataset Composition}
\label{tab:dataset_composition}
\resizebox{.95\columnwidth}{!}{
\begin{tabular}{p{2cm}p{2cm}p{3cm}}
\toprule
\textbf{Source} & \textbf{Instances} & \textbf{Annotation} \\
\midrule
PHATE & 450 & Wrote CNs and labeled their types \\
MT-CONAN & 150 & Translated CNs and labeled their types\\
\bottomrule
\end{tabular}}
\end{table}

\subsubsection{Data Sourcing}
\label{subsubsec:data_sourcing_one}
To address the scarcity of Persian hate speech data, we sourced 450 hate speech instances from the PHATE dataset \citep{delbari2024spanning}, selected to represent diverse expressions of hate across six categories. However, PHATE provided only 20 and 30 unique and acceptable instances for Gender and Racial categories, respectively, necessitating supplementation to achieve 100 instances per category. To ensure cultural relevance for Persian contexts, we carefully selected 150 instances from MultiTarget-CONAN \citep{fanton2021human} and HateXplain \citep{mathew2021hatexplain}. From MultiTarget-CONAN, we chose 80 instances targeting Women for the Gender category, reflecting prevalent Persian socio-cultural issues like gender-based discrimination (e.g., claims that women are unfit for politics, countered with arguments for gender equality). For the Racial category, we selected 70 instances targeting People of Color, ensuring alignment with ethnic tensions in Persian discourse (e.g., derogatory remarks about Arabs, countered with calls for humanity). From HateXplain, we selected instances targeting groups like Chinese and Arabs, which frequently appear in Persian online discourse due to geopolitical and cultural interactions. Full examples, with Persian and English translations to illustrate cultural fit are shown in Figure \ref{fig:translated-examples} (see Appendix E for details). All instances from MultiTarget-CONAN and HateXplain were translated into Persian using Google Translate and manually post-edited by native Persian annotators to ensure linguistic precision, grammatical accuracy, and cultural congruence, aligning with Persian-specific target categories. This process yielded a balanced dataset of 600 instances, ensuring that translated samples reflect the socio-cultural nuances of Persian online discourse. To mitigate potential translation bias, our post-editing process functioned as content localization rather than literal translation. We replaced Western-centric hate tropes with concepts relevant to Persian socio-cultural dynamics (e.g., shifting general racial slurs to specific regional ethnic tensions). The success of this approach is evidenced by our human evaluation results (Section titled ``Human Evaluation'' and Appendix I), where translated-and-adapted samples achieved strong pair-level human evaluation scores that were broadly comparable to native Persian instances, especially in fluency, semantic clarity, and tone appropriateness.
\begin{figure*}[ht]
  \centering
  \includegraphics[width=0.97\textwidth]{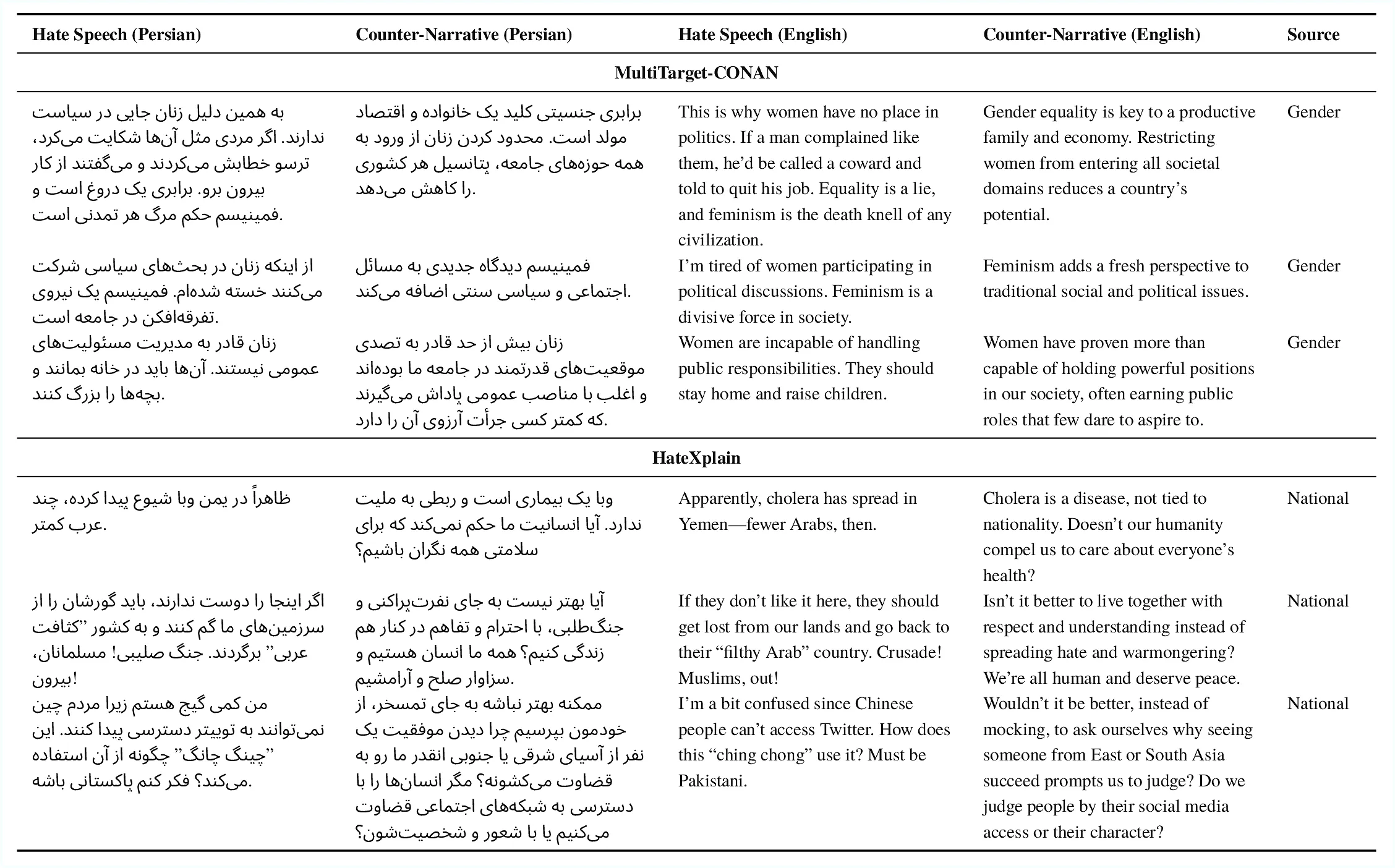} 
  \caption{Examples of Culturally Relevant Hate Speech and Counter-Narratives from MT-CONAN and HateXplain with English Translations}
  \label{fig:translated-examples}
\end{figure*}

\subsubsection{Counter-Narrative Annotation}
\label{subsubsec:counter_narrative_annotation}

Securing multiple annotations for a low-resource task like Persian counter-narrative generation is inherently challenging due to the limited availability of individuals who are both fluent in Persian and experienced in hate speech research. This constraint is also evident in prior work, such as the IndicCONAN dataset \citep{sahoo2024indicconan}, which relied on only two annotators. In our case, annotation was conducted by three native Persian speakers with prior research experience in hate speech and counter speech (See Appendix F for detailed annotator information). For the 600 human-generated hate speech–counter-narrative pairs, the annotation process began with a shared batch of 80 samples, which all three annotators worked on independently. They then discussed their responses collaboratively to identify and correct mistakes, refining their annotation strategy and aligning their interpretations to ensure high-quality outputs. After this initial phase, the remaining samples were distributed among the annotators for individual annotation. For each hate speech instance, annotators generated a single counter-narrative, guided by comprehensive guidelines adapted from PEN America and Get The Trolls Out ( \url{https://tinyurl.com/ycxct6wh}). These guidelines outlined strategies for crafting respectful, empathetic, and effective counter-narratives, emphasizing the use of credible information and appropriate tone. Counter-narrative type selection (e.g., Positive Response, Denouncing, Fact-based, as defined in Appendix B) was informed by these guidelines and the context of the hate speech, with a focus on producing impactful and culturally resonant responses. While no rigid template was used—preserving cultural nuance—the use of guidelines and initial collaborative annotation helped minimize subjectivity. For example, a gender-based hate speech suggesting that women are untrustworthy was countered with: “Labeling all women this way is entirely unjustified.” Throughout the annotation process, we aimed to balance diversity in counter-narrative types—avoiding over-reliance on strategies like Denouncing—while maintaining cultural sensitivity. Each counter-narrative was labeled with its type, in accordance with the framework outlined in the Counter-Narrative Approaches section.

\subsubsection{MT-CONAN Integration}
\label{subsubsec:mt_conan_integration}

To enhance dataset diversity, the 150 MT-CONAN instances were paired with their original counter-narratives, translated into Persian using Google Translate and post-edited by annotators for fluency and cultural relevance. Annotators labeled each translated counter-narrative with its corresponding type, ensuring consistency with the PHATE annotations. This process, focused on the Gender (80 women-targeted instances) and Racial (70 poc targeted instances) categories, introduced varied counter-narrative styles while maintaining dataset coherence. Quality control during post-editing addressed challenges like preserving the original counter-narratives’ intent in Persian contexts. The combined dataset of 600 hate speech--counter-narrative pairs provides a robust foundation for Stage Two’s model training.

\subsection{Stage Two: Automated Counter-Narrative Generation}
\label{subsec:stage_two_generation}

The second stage focused on expanding the dataset by leveraging LLMs for automated counter-narrative generation. This stage aimed to produce an additional 500 hate speech--counter-narrative pairs across five target categories, utilizing the manually curated Stage One data as few-shot examples to guide the LLMs towards generating high-quality, contextually relevant responses in Persian. The process involved sourcing new hate speech instances, retrieving relevant examples from Stage One, prompting the LLMs for generation, and performing manual curation and quality control on the generated outputs.

\subsubsection{Expansion of Dataset Sourcing}
\label{subsubsec:data_sourcing_two}

To provide the LLMs with new hate speech instances for generating counter-narratives, we sourced 100 new hate speech instances for each of the five target categories: Religious, Political, Gender, Racial, and National. Instances for the Religious and Political categories were collected from the PHATE dataset \citep{delbari2024spanning}. For the Gender (specifically targeting women) and Racial categories, we selected instances from the MT-CONAN dataset \citep{fanton2021human}. Instances for the National category were sourced from the HateXplain dataset \citep{mathew2021hatexplain}. We carefully selected these instances to ensure their relevance and appropriateness for generating counter-narrative responses in the Persian context, as we discussed in the Data Sourcing section. The Occupational target group was excluded in this stage due to difficulty in identifying a sufficient number of high-quality instances suitable for counter-narrative generation that aligned with the dataset's goals. For the English-language sources (MT-CONAN and HateXplain), the selected hate speech instances were translated into Persian using Google Translate and subsequently reviewed and manually edited by annotators to ensure linguistic accuracy, fluency, and cultural coherence within the Persian online discourse.

\subsubsection{Retrieval of Relevant Examples}
\label{subsubsec:example_retrieval}
To implement the few-shot prompting strategy effectively, we developed a method to retrieve the most semantically similar hate speech--counter-narrative pairs from the Stage One dataset (600 pairs) for each new hate speech instance collected in Stage Two. We utilized a pre-trained SentenceTransformer model, specifically \texttt{`paraphrase-multilingual-mpnet-base-v2'}, known for its effectiveness in generating multilingual sentence embeddings. Each new hate speech instance from Stage Two was embedded using this model, as were all hate speech instances from the Stage One dataset. Cosine similarity was then calculated between the embedding of the new instance and the embeddings of all Stage One instances. For each new hate speech instance, the top 10 most similar pairs from Stage One, along with their similarity scores and metadata (original hate speech, counter-narrative, target group, counter type(s)), were retrieved. These retrieved pairs served as examples in the LLM prompts, providing contextual guidance on the desired style and content of effective Persian counter-narratives for similar hate speech. The retrieved pairs were manually reviewed before inclusion in the prompts to select the most relevant and high-quality examples.

\subsubsection{LLM-Based Counter-Narrative Generation}
\label{subsubsec:llm_generation}
The automated generation of counter-narratives was performed using four LLMs: GPT-4o, Gemini 2.0 Flash, Claude 3.7 Sonnet, and Qwen3-235B-A22B, selected for their strong multilingual capabilities. A few-shot prompting approach was employed, where the input to the LLM included the new hate speech instance requiring a response, along with the corresponding hate speech-counter-narrative pairs retrieved from Stage One, and explicit instructions regarding the desired output format and content. The prompt instructed the models to generate a short counter-narrative in Persian, ideally between 1 and 2 lines and under 50 words. Furthermore, the prompt encouraged the LLMs to utilize a combination of two counter-narrative types from the predefined list (Positive Response, Counter Questions, Denouncing, Fact-based, Warning of Consequences, Contradiction and Hypocrisy), or to select the single most effective type based on the provided examples and type definitions. Across the five target categories, each of the four LLMs generated 25 counter-narratives per category. This process resulted in a total of 4 LLMs $\times$ 5 categories $\times$ 25 instances/category = 500 automatically generated counter-narratives.

\begin{tcolorbox}[
    title=\textbf{Persian Counter-Narrative Generation Prompt},
    colback=blue!5!white,
    colframe=blue!75!black,
    fonttitle=\bfseries,
    width=0.46\textwidth, 
    boxrule=0.8pt,
    arc=4pt,
    enlarge left by=0mm, 
    enlarge right by=0mm, 
    left=4pt,
    right=4pt,
    top=6pt,
    bottom=6pt,
]
\small
\label{tab:prompt}
Generate a counter-narrative response in Persian to the hate speech: \textbf{\{HATE\_SPEECH\}}. The response must:

\begin{itemize}[leftmargin=1.5em]
    \item Be a short paragraph (1–2 lines, under 50 words).
    \item Be in Persian.
    \item Use a mix of \textbf{\{TYPE 1\}} and \textbf{\{TYPE 2\}} counter-narrative types or select the most effective type from the following, based on these sample counter-narrative responses:
    \begin{itemize}
        \item \textbf{Positive Response}: Statements presenting an optimistic, constructive, or supportive viewpoint to promote understanding, empathy, or positive change.
        \item \textbf{Counter Questions}: Thought-provoking questions that challenge underlying assumptions, biases, or implications to stimulate critical thinking.
        \item \textbf{Denouncing}: Statements openly condemning harmful ideas or actions, highlighting negative effects or ethical concerns.
        \item \textbf{Fact-based}: Statements providing factual evidence to correct misperceptions or misinformation.
        \item \textbf{Warning of Consequences}: Statements informing of potential negative outcomes of a viewpoint or action.
        \item \textbf{Contradiction and Hypocrisy}: Statements pointing out inconsistencies or hypocrisy in the original statement.
    \end{itemize}
\end{itemize}

\textbf{Examples:}
\begin{itemize}
    \item \{Hate Speech1\}, \{Counter-Narrative1\}, \{Target Group\}, \{Counter Type\}
    \item \{Hate Speech2\}, \{Counter-Narrative2\}, \{Target Group\}, \{Counter Type\}
    \item \{Hate Speech n\}, \{Counter-Narrative n\}, \{Target Group\}, \{Counter Type\}
\end{itemize}

\textbf{Do not include explanations, notes, or multiple responses—only the counter-narrative paragraph.}
\end{tcolorbox}

\subsubsection{Manual Curation and Quality Control}
\label{subsubsec:manual_curation}
Although the counter-narratives were generated using advanced LLMs, a crucial manual curation and quality control step was performed to ensure the high quality and usability of the expanded dataset. All 500 automatically generated counter-narratives were meticulously reviewed by human annotators. This review process focused on several key aspects: evaluating the fluency and grammatical correctness of the Persian text, assessing the cultural appropriateness, verifying adherence to the specified length constraints (under 50 words), and confirming that the generated response effectively employed one or a combination of the intended counter-narrative types. Annotators made necessary edits to correct errors, improve clarity, and ensure that each generated counter-narrative met the dataset's quality standards. This manual refinement step was essential for producing a reliable and high-quality dataset extension for training and evaluating counter-narrative models in Persian.

\subsection{FAIR Compliance} The dataset has been developed in accordance with the FAIR principles to ensure that it is Findable, Accessible, Interoperable, and Reusable. The dataset is assigned a persistent DOI (10.5281/zenodo.18266930),  making  it  easily  findable. The dataset is hosted on a data-sharing platform  that  allows  for authorized  access  to  researchers, ensuring  that  it  can  be  retrieved  using  standard  protocols. The data is stored in standard CSV format, which is widely used and compatible with common data analysis tools. This promotes interoperability with various software applications. The dataset is shared under Creative Commons Attribution 4.0 International, which allows for reuse in academic research.

\subsection{Potential Research Applications and Usage}
This dataset enables the development and evaluation of automated systems for Persian counter-narrative detection, a task that has been largely unexplored due to the lack of annotated resources. It can be used to train and benchmark machine learning models for counter-narrative classification, stance analysis, and response generation. The dataset also supports sociolinguistic research on how Persian-speaking communities respond to online hate, facilitating cross-cultural comparisons and multilingual transfer learning. In practical settings, it can aid the creation of moderation tools that recommend constructive responses, prioritize harmful content, and promote healthier online interactions. Additionally, the resource provides a benchmark for assessing the cultural sensitivity and safety of large language models in Persian. Overall, it contributes to advancing computational methods and real-world interventions for mitigating online hate in Persian digital spaces.

\section{Data Statistics}
\label{sec:data_statistics}
This section presents key statistics describing the composition and characteristics of the \textbf{ParsCN} dataset. 

\subsection{Dataset Composition}
\label{subsec:composition_and_lengths}
Table~\ref{tab:composition_and_lengths} presents a combined view of the dataset's composition by target group and the average word counts for both hate speech instances and their corresponding counter-narratives within each group. These statistics provide insight into the distribution of pairs across the six defined target groups and the typical verbosity of hate speech and counter-narratives. The dataset comprises a total of 1100 hate speech--counter-narrative pairs. 
We observe variations in average word counts across target groups. The Religious group has the longest average counter-narrative word count (32.04 words), while the Political group has the shortest (23.98 words). We limited counter-narratives to under 50 words (typically 1-3 sentences). This choice is empirically motivated by social media consumption patterns on platforms like Twitter/X, Instagram etc., where concise responses demonstrate significantly higher readability, mobile engagement, and likelihood of dissemination compared to long-form rebuttals.

\begin{table}[h]
\centering
\small
\caption{Dataset Composition and Average Word Counts (HS/CN) by Target Group}
\label{tab:composition_and_lengths}
\resizebox{.95\columnwidth}{!}{
\begin{tabular}{p{2cm} p{1.2cm} p{1.5cm} p{1.8cm}}
\toprule
\textbf{Target Group} & \textbf{Number of Pairs} & \textbf{Average HS Word Count} & \textbf{Average CN Word Count} \\
\midrule
Gender & 200 & 18.96 & 26.18 \\
Political & 200 & 37.48 & 23.98 \\
National & 200 & 31.85 & 24.55 \\
Racial & 200 & 17.07 & 26.28 \\
Religious & 200 & 35.77 & 32.04 \\
Occupational & 100 & 36.16 & 25.01 \\
\midrule
\textbf{Overall Avg} & \textbf{1100} & \textbf{29.55} & \textbf{26.34} \\
\bottomrule
\end{tabular}}
\end{table}

\subsection{Distribution of Counter-Narrative Types}
\label{subsec:counter_type_distribution}

The dataset's counter-narratives are annotated with six primary types, often appearing in combination. Among these, Positive Response is the most frequently occurring type, appearing in 547 counter-narratives, reflecting a strong preference for constructive, empathetic, or inclusive strategies. Denouncing follows with 400 occurrences, indicating that directly condemning hate speech is also a common approach. Counter Questions appear 323 times, suggesting moderate use of reflective or rhetorical challenges. Less frequent are Fact-based counter-narratives (221), Warning of Consequences (185), and Contradiction (168), which may point to the difficulty of countering emotionally charged hate speech with logic, or to the specific nature of hate speech in the dataset. Overall, the prevalence of Positive Response highlights the dataset’s emphasis on fostering positive discourse as a central countering mechanism. This imbalance is partly intentional and partly data-driven. Our annotation guidelines encouraged constructive, respectful, and non-escalatory responses, which made Positive Response broadly applicable across many hate speech scenarios. By contrast, strategies such as Fact-based, Warning of Consequences, and Contradiction often require stronger contextual, factual, or rhetorical grounding and were therefore less universally suitable. We also acknowledge that some portion of the observed distribution may reflect annotator preference, despite our best effort to maintain diversity across response types. We view this distribution not as a flaw, but as an informative characteristic of how counter-narrative is naturally operationalized in Persian online contexts.

\section{Human Evaluation}
\label{sec:human_evaluation}

To assess the quality of the generated counter-narratives from different sources (LLMs, MT-CONAN, and human writers), we conducted a human evaluation. Two expert annotators evaluated a subset of the generated counter-narratives based on several key metrics.

\subsection{Evaluation Metrics}
\label{subsec:human_evaluation_metrics}

We used the following metrics to evaluate the counter-narratives:\\
\textbf{Relevance}: Measures how well the counter-narrative addresses the original hate speech content and context. \\
\textbf{Effectiveness}: Evaluates the persuasive power or potential impact of the counter-narrative in mitigating the hateful intent or sentiment of the original hate speech. \\
\textbf{Fluency}: Assesses the grammatical correctness and readability of the generated text in Persian. \\
\textbf{Tone Appropriateness}: Evaluates whether the counter-narrative uses a respectful, empathetic, and appropriate tone without escalating or causing further harm.

\subsection{Inter-Annotator Agreement Scores}
\label{subsec:inter_annotator_agreement}
To ensure the reliability and consistency of the manual annotation process, particularly for counter-narrative quality assessment metrics defined above, we calculated Cohen's Kappa scores between the two independent expert annotators. As the original human evaluation scores were assigned on a continuous-like scale (1 to 5 with 0.5 increments), and Cohen's Kappa is typically used for categorical data, we categorized the scores into three distinct levels: Low' ([1, 1.5, 2]), Medium' ([2.5, 3, 3.5]), and `High' ([4, 4.5, 5]). This categorization, consistent with prior NLP annotation studies \citep{Landis1977TheMO}, avoids overlap and ensures clarity in distinguishing score ranges. Kappa was then computed for each annotation criterion on a subset of the annotated data used in the human evaluation. The resulting agreement scores demonstrate substantial inter-annotator agreement: 0.724 for Relevance, 0.692 for Effectiveness, 0.655 for Fluency, and 0.770 for Tone Appropriateness. These results indicate a high level of consistency and reliability.

\subsection{Evaluation Results}
\label{subsec:human_evaluation_results}

Table~\ref{tab:human_evaluation_scores} presents the average human evaluation scores for each source across the four metrics, as rated by two independent annotators (A1 and A2), along with their combined average. Scores were rated on a scale from 1-5.
We observe that the Human-written counter-narratives receive the highest average scores across all metrics, confirming the effectiveness of human intuition, context-awareness, and nuanced language generation in sensitive scenarios. Among the models, GPT-4o and Claude closely follow in performance, especially in relevance and tone appropriateness, showing strong generalization and emotional calibration. Gemini scores well in fluency, while Qwen leads in relevance but slightly lags in fluency. MT-CONAN, while competitive in fluency, exhibits lower effectiveness, indicating a potential trade-off between linguistic smoothness and persuasive impact. 

\begin{table*}[ht!]
\centering
\small
\caption{Human Evaluation Scores by Source and Annotator Average}
\label{tab:human_evaluation_scores}
\resizebox{.9\textwidth}{!}{
\begin{tabularx}{\textwidth}{l *{12}{>{\centering\arraybackslash}X}}
\toprule
\textbf{Source} & \multicolumn{3}{c}{\textbf{Relevance}} & \multicolumn{3}{c}{\textbf{Effectiveness}} & \multicolumn{3}{c}{\textbf{Fluency}} & \multicolumn{3}{c}{\textbf{Tone Appropriateness}} \\
\cmidrule(lr){2-4} \cmidrule(lr){5-7} \cmidrule(lr){8-10} \cmidrule(lr){11-13}
& A1 & A2 & Avg & A1 & A2 & Avg & A1 & A2 & Avg & A1 & A2 & Avg \\
\midrule
Claude          & 4.175 & 4.088 & 4.132 & \textbf{4.050} & 4.138 & 4.094 & 4.850 & 4.800 & 4.825 & 4.850 & 4.488 & 4.669 \\
GPT-4o          & 4.162 & 4.125 & 4.144 & 4.037 & 4.188 & 4.113 & 4.750 & 4.688 & 4.719 & 4.825 & 4.562 & 4.693 \\
Gemini          & 4.225 & 4.000 & 4.112 & 3.925 & 3.950 & 3.938 & 4.888 & 4.725 & 4.807 & 4.700 & 4.412 & 4.556 \\
Human           & 4.150 & 4.312 & \textbf{4.231} & 4.025 & 4.388 & \textbf{4.206} & 4.950 & 4.888 & \textbf{4.919} & 4.962 & 4.612 & \textbf{4.787} \\
MT-CONAN            & 4.162 & 3.888 & 4.025 & 3.775 & 3.738 & 3.756 & 4.712 & 4.800 & 4.756 & 4.588 & 4.425 & 4.507 \\
Qwen            & \textbf{4.250} & 4.050 & 4.150 & 3.988 & 4.025 & 4.006 & 4.638 & 4.362 & 4.500 & 4.850 & 4.625 & 4.738 \\
\bottomrule
\end{tabularx}}
\end{table*}

\section{Automatic Evaluation}
\label{sec:automatic_evaluation}

To complement human evaluation and provide quantitative insights into the quality and characteristics of the generated counter-narratives, we conducted an automatic evaluation using several widely adopted metrics. This evaluation assessed the fluency, semantic similarity, lexical diversity, and toxicity of counter-narratives generated by different LLMs, translated from MT-CONAN, and written by native speakers.

\subsection{Evaluation Metrics and Results}
We employed several automatic evaluation metrics to assess different aspects of the generated counter-narratives.  Table~\ref{tab:combined_evaluation_metrics} presents a combined view of the results for Perplexity \citep{jelinek1977perplexity}, Semantic Similarity (BERTScore) \citep{zhang2019bertscore}, Lexical Diversity (Distinct-n) \citep{li2015diversity}, and Toxicity \citep{imani2023glot500} across different data sources (LLMs, MT-CONAN, and Human). 

\begin{table*}[ht!]
\centering
\small
\caption{Automatic Evaluation Metrics by Source. 
(P: Precision, R: Recall, F1: F1 Score)}
\label{tab:combined_evaluation_metrics}
\resizebox{.9\textwidth}{!}{
\begin{tabularx}{\textwidth}{l r *{3}{>{\centering\arraybackslash}X} *{2}{>{\centering\arraybackslash}X} c}
\toprule
& \textbf{Perplexity} & \multicolumn{3}{c}{\textbf{BERTScore}} & \multicolumn{2}{c}{\textbf{Distinct-n}} & \textbf{Toxicity} \\
\cmidrule(lr){3-5} \cmidrule(lr){6-7}
\textbf{Source} & & P & R & F1 & Distinct-1 & Distinct-2 & Score \\
\midrule
Gemini & \textbf{39.15} & 0.660 & 0.666 & 0.662 & 0.303 & 0.754 & \textbf{0.037} \\
Claude & 49.37 & 0.659 & 0.667 & 0.663 & 0.317 & 0.782 & 0.053 \\
Qwen & 61.70 & 0.666 & 0.672 & 0.669 & 0.302 & 0.740 & 0.078 \\
GPT-4o & 99.60 & 0.664 & 0.669 & 0.666 & \textbf{0.410} & 0.873 & 0.088 \\
MT-CONAN & 100.42 & \textbf{0.692} & \textbf{0.728} & \textbf{0.709} & 0.405 & \textbf{0.892} & 0.106 \\
Human & \textbf{158.45} & 0.643 & 0.628 & 0.635 & \textbf{0.384} & \textbf{0.879} & 0.061 \\
\bottomrule
\end{tabularx}}
\end{table*}


\noindent \textbf{Perplexity:} LLM-generated CNs generally exhibit lower perplexity than MT-CONAN translations and human samples, suggesting higher fluency and consistency with typical language patterns learned by the models. Human-written CNs show the highest perplexity (158.45). While numerically higher perplexity might seem undesirable, in this context, it is a valuable indicator of native speaker written data's linguistic richness, creativity, and inclusion of more complex or nuanced phrasing that is less predictable by standard language models. This inherent variability is crucial for training models that can generate diverse and human-like responses.

\noindent \textbf{Semantic Similarity (BERTScore):} MT-CONAN translated CNs achieve the highest semantic similarity scores, indicating strong alignment with their hate speech text. LLM-generated CNs follow closely, demonstrating their capability to produce semantically relevant counter-narratives. Human-written CNs show slightly lower scores and reflect the more varied and interpretive ways humans respond, adding valuable layers of meaning and context.

\noindent \textbf{Lexical Diversity (Distinct-n):} The MT-CONAN responses (0.892 Distinct-2) and human-written CNs (0.384 Distinct-1, 0.879 Distinct-2) exhibit high lexical diversity, reflecting a broad range of vocabulary and phrasing. LLMs, particularly GPT-4o (0.410 Distinct-1), also demonstrate competitive lexical diversity, balancing fluency with varied language use. 

\noindent \textbf{Toxicity:} We estimated toxicity using the Glot500 classifier \citep{imani2023glot500}. Lower scores indicate less toxicity. All sources produce predominantly non-toxic counter-narratives. LLM-generated responses, especially from Gemini and Claude, show extremely low toxicity. Human responses maintain a moderate level, balancing assertiveness with civility. MT-CONAN samples have higher toxicity, potentially influenced by translation nuances or the original source's tone.

The automatic evaluation metrics collectively highlight the strengths of the ParsCN dataset, which integrates data from multiple sources. LLMs contribute fluency and semantic relevance, while MT-CONAN data brings high semantic similarity and lexical diversity. Human-authored counter-narratives offer invaluable linguistic richness and realistic variation. The low toxicity scores across all sources confirm the dataset's focus on generating appropriate and non-harmful counter-narrative. This blend of characteristics makes ParsCN a robust and diverse resource for developing and evaluating counter-narrative generation models in Persian.

\section{Baseline Model Evaluation}
To underscore the necessity of ParsCN, we evaluated two baseline models, mBART and PersianMind, on a representative subset of 125 Persian hate speech samples from the ParsCN dataset. We selected this subset to balance diversity across target groups with the practical cost of manual inspection and human evaluation. The experiment setup and results are presented in Appendix D (Tables \ref{tab:standard_metrics}, \ref{tab:quality_safety}, and \ref{tab:human_metrics}). The models were assessed using automated metrics (BLEU, ROUGE-L, METEOR, BERTScore F1, perplexity, toxicity) and human evaluations (relevance, effectiveness, tone appropriateness, fluency, cultural relevance). mBART, fine-tuned on English MT-CONAN data, showed slightly better lexical similarity (BLEU: 0.0096, ROUGE-L: 0.0855, METEOR: 0.0892) but had high toxicity (0.2640 vs. 0.0163 for gold references) and low human scores (e.g., relevance: 2.35, cultural relevance: 1.95 vs. 3.97 and 3.68 for gold references). PersianMind, a monolingual Persian model, scored near-zero on BLEU and ROUGE-L, with high perplexity (316.3498 vs. 116.1709 for gold references), indicating poor fluency. Its human evaluation scores (e.g., effectiveness: 2.02 vs. 4.04 for gold references) further highlight limitations in producing relevant and culturally aligned responses. Despite reasonable BERTScore F1 scores (0.8305–0.8464), both models generated outputs that were generic and lacked the specificity and cultural nuance present in ParsCN’s human-annotated responses. These results reveal significant gaps in existing models’ ability to generate precise, fluent, safe, and culturally appropriate Persian counter-narratives—primarily due to the lack of Persian-specific training data. mBART’s reliance on English data fails to capture Persian linguistic and cultural complexities, while PersianMind lacks fine-tuning on domain-specific counter-narrative tasks. ParsCN addresses this critical gap by providing a culturally tailored, high-quality annotated dataset to enable the development of effective counter-narrative generation models for Persian-speaking communities.

\section{Conclusions and Future Work}
\label{sec:conclusions}
We introduce \textbf{ParsCN}, the first large-scale, carefully curated corpus for Persian counter-narrative generation. The dataset comprises 1,100 hate speech–counter-narrative pairs spanning six target groups and six countering strategies. Built through a multi-stage pipeline that combines expert annotation, strategic translation, and few-shot prompting of LLMs—each step followed by rigorous human curation—\textbf{ParsCN} offers a rich, balanced, and culturally grounded resource for both training and evaluating counter-narrative systems in a low-resource language context.

\textbf{ParsCN}’s fine-grained annotations and balanced structure make it an immediate test bed for advancing counter-narrative generation and evaluation in Persian. Beyond its intrinsic value, the dataset illustrates how a human-in-the-loop, LLM-assisted approach can efficiently generate high-quality, culturally appropriate resources at a fraction of the usual cost—offering a scalable and cost-effective template for supporting counter-narrative research in other low-resource languages.

Looking ahead, several promising directions can extend and build upon \textbf{ParsCN}. The dataset can be enriched by expanding its coverage to include additional target groups (e.g., age, disability), adopting more nuanced or culturally specific counter-narrative strategies, and incorporating fine-grained annotations—such as the severity of hate speech or the perceived effectiveness of responses. Potential future work may include designing and experimenting with automatic counter-narrative generation models, e.g., fine-tuning LLMs, exploring retrieval-augmented approaches, and examining strategy-optimized architectures. Future work should also move beyond offline dataset construction and evaluation to study how counter-narratives perform in real online environments. In particular, an important next step is to examine whether generated responses can meaningfully influence user attitudes, reduce hostility, or support de-escalation in practice. In this sense, \textbf{ParsCN} should be viewed as a foundational benchmark for Persian counter-narrative generation and a basis for future user-centered and deployment-oriented evaluation. Moreover, the dataset provides a foundation for cross-lingual research, enabling the transfer of counter-narrative capabilities to other languages and supporting the creation of robust multilingual frameworks for combating online hate.

\bibliography{references}
\subsection{Paper Checklist}

\begin{enumerate}

\item For most authors...
\begin{enumerate}
    \item  Would answering this research question advance science without violating social contracts, such as violating privacy norms, perpetuating unfair profiling, exacerbating the socio-economic divide, or implying disrespect to societies or cultures?
    Yes
  \item Do your main claims in the abstract and introduction accurately reflect the paper's contributions and scope? Yes
   \item Do you clarify how the proposed methodological approach is appropriate for the claims made? 
  Yes
   \item Do you clarify what are possible artifacts in the data used, given population-specific distributions?
   Yes
  \item Did you describe the limitations of your work?
   Yes, see Appendix J
  \item Did you discuss any potential negative societal impacts of your work?
    Yes, see Appendix G
      \item Did you discuss any potential misuse of your work?
    NA
    \item Did you describe steps taken to prevent or mitigate potential negative outcomes of the research, such as data and model documentation, data anonymization, responsible release, access control, and the reproducibility of findings?
    Yes
  \item Have you read the ethics review guidelines and ensured that your paper conforms to them?
    Yes
\end{enumerate}

\item Additionally, if your study involves hypotheses testing...
\begin{enumerate}
  \item Did you clearly state the assumptions underlying all theoretical results?
    NA
  \item Have you provided justifications for all theoretical results?
    NA
  \item Did you discuss competing hypotheses or theories that might challenge or complement your theoretical results?
   NA
  \item Have you considered alternative mechanisms or explanations that might account for the same outcomes observed in your study?
    NA
  \item Did you address potential biases or limitations in your theoretical framework?
   NA
  \item Have you related your theoretical results to the existing literature in social science?
   NA
  \item Did you discuss the implications of your theoretical results for policy, practice, or further research in the social science domain?
    NA
\end{enumerate}

\item Additionally, if you are including theoretical proofs...
\begin{enumerate}
  \item Did you state the full set of assumptions of all theoretical results?
    NA
	\item Did you include complete proofs of all theoretical results?
    NA
\end{enumerate}

\item Additionally, if you ran machine learning experiments...
\begin{enumerate}
  \item Did you include the code, data, and instructions needed to reproduce the main experimental results (either in the supplemental material or as a URL)?
    Yes, see section Appendix A
  \item Did you specify all the training details (e.g., data splits, hyperparameters, how they were chosen)?
    NA
     \item Did you report error bars (e.g., with respect to the random seed after running experiments multiple times)?
    NA
	\item Did you include the total amount of compute and the type of resources used (e.g., type of GPUs, internal cluster, or cloud provider)?
    No
     \item Do you justify how the proposed evaluation is sufficient and appropriate to the claims made? 
    Yes
     \item Do you discuss what is ``the cost`` of misclassification and fault (in)tolerance?
    NA
  
\end{enumerate}

\item Additionally, if you are using existing assets (e.g., code, data, models) or curating/releasing new assets, \textbf{without compromising anonymity}...
\begin{enumerate}
  \item If your work uses existing assets, did you cite the creators?
  Yes
  \item Did you mention the license of the assets?
  Yes
  \item Did you include any new assets in the supplemental material or as a URL?
   Yes
  \item Did you discuss whether and how consent was obtained from people whose data you're using/curating?
   NA
  \item Did you discuss whether the data you are using/curating contains personally identifiable information or offensive content?
   Yes, the dataset does not contain PII, however, since it includes hate speeches, offensive content is included.
\item If you are curating or releasing new datasets, did you discuss how you intend to make your datasets FAIR (see \citet{fair})?
Yes, see section titled ``FAIR Compliance''
\item If you are curating or releasing new datasets, did you create a Datasheet for the Dataset (see \citet{gebru2021datasheets})? 
Yes
\end{enumerate}

\item Additionally, if you used crowdsourcing or conducted research with human subjects, \textbf{without compromising anonymity}...
\begin{enumerate}
  \item Did you include the full text of instructions given to participants and screenshots?
   Yes
  \item Did you describe any potential participant risks, with mentions of Institutional Review Board (IRB) approvals?
    The study is exempted by the IRB
  \item Did you include the estimated hourly wage paid to participants and the total amount spent on participant compensation?
    The annotators voluntarily contributed to the work. No hourly wage was paid.
   \item Did you discuss how data is stored, shared, and deidentified?
   Yes
\end{enumerate}

\end{enumerate}

\appendix

\section{Appendix A: Data and Code Availability}

Both the ParsCN dataset and the full codebase used for automatic evaluation (BERTScore, BLEU, ROUGE, METEOR, Distinct‑n, Perplexity, Toxicity) as well as for baseline model evaluation (mBART and PersianMind) are available on Github at: \texttt{\url{https://github.com/zahrasafdari/ParsCN-Dataset}}

The repository includes:
\begin{itemize}
  \item Evaluation scripts for all automatic metrics described in the paper.
  \item Baseline model training and inference code.
  \item Pre‑processing routines for dataset tokenization and formatting.
  \item The proposed ParsCN dataset that contains hate speech-counter-narrative pairs along with target groups and counter-narrative types.
\end{itemize}

\textbf{License:}  The materials are publicly released under a CC BY 4.0 license.


\section{Appendix B: Detailed Counter-Narrative Types' Definition}
\label{app:type_definitions}

The counter-narrative types used in the dataset are described below. These definitions are derived from \citet{sahoo2024indicconan}:

\begin{itemize}
\item \textit{Positive Response:} These counter-narratives counter hostility with inclusive, supportive statements that encourage empathy and unity. For example, in response to gender-based hate, they might advocate mutual respect in relationships to promote harmony. This approach seeks to reframe divisive rhetoric into constructive dialogue.

\item  \textit{Counter questions:} This type uses probing questions to challenge the biases or assumptions in hate speech, urging reflection. For instance, questioning claims about Sunni minorities’ ambitions can prompt reconsideration of stereotypes, fostering critical thinking.

\item  \textit{Denouncing:} These statements firmly reject harmful rhetoric, condemning its ethical or social damage. For example, denouncing slurs against religious groups as divisive aims to curb their spread by highlighting their harm.

\item  \textit{Fact-based:} This approach refutes misinformation with credible evidence, enhancing discourse accuracy. For instance, citing statistics to debunk myths about women’s driving in Iran corrects false narratives and builds trust in factual dialogue.

\item  \textit{Warning of consequences:} These counter-narratives caution against the adverse effects of hate speech, such as social division from targeting Afghan immigrants. By emphasizing risks, they encourage more responsible perspectives.

\item  \textit{Contradiction:} This type exposes inconsistencies in hate speech, such as criticizing women’s autonomy while ignoring men’s. By highlighting logical flaws, it undermines the credibility of harmful claims.
\end{itemize}



\section{Appendix C: Comparison of Low-Resource Counter-Narrative Datasets}

This appendix provides a detailed comparison of ParsCN with other low-resource counter-narrative (CN) datasets to highlight its novel contributions in scope and methodology. Table \ref{tab:dataset_comparison} summarizes key differences in language, data sourcing, annotation approach, and scalability.

\begin{table}[h]
\centering
\small
\caption{Comparison of Low-Resource CN Datasets}
\label{tab:dataset_comparison}
\setlength{\tabcolsep}{3pt} 
\renewcommand{\arraystretch}{1.1}
\resizebox{0.9\columnwidth}{!}{%
\begin{tabular}{l@{\hskip 
8pt}p{1.5cm}@{\hskip 8pt}p{2cm}@{\hskip 8pt}p{2.5cm}@{\hskip 8pt}p{1.5cm}}
\toprule
\textbf{Dataset} & \textbf{Lang.} & \textbf{Sourcing} & \textbf{Annotation} & \textbf{Scal.} \\
\midrule
CONAN & Eng., Fr., It. & Nichesourcing & Expert-based & Limited \\
IndicCONAN & Indian languages. & Crowdsourcing & Human-only & Moderate \\
PANDA & Chinese & LLM-as-a-Judge & Automated & High \\
CONAN-EUS & Basque, Spanish & Translation & Post-edited & Moderate \\
\textbf{ParsCN} & \textbf{Persian} & \textbf{Hybrid (PHATE, translated)} & \textbf{Human+LLM}, \textbf{culturally tailored} & \textbf{High} \\
\bottomrule
\end{tabular}%
}
\end{table}

\section*{Appendix D: Baseline Model Evaluation Details}

This appendix provides the detailed experiment setup and results for evaluating baseline models (mBART and PersianMind) on generating counter-narratives for 125 Persian hate speech samples from the ParsCN dataset.

\subsection*{Experiment Setup}
We assessed two models on generating counter-narrative responses for 125 diverse hate speech samples from the ParsCN dataset:

\begin{itemize}
\item \textbf{mBART}: A multilingual sequence-to-sequence model fine-tuned on the English MT-CONAN dataset. Persian hate speech samples were translated into English by human translators with post-editing, and mBART generated English counter-narratives, compared against gold-standard English-translated Persian counter-narratives.
\item \textbf{PersianMind} (universitytehran/PersianMind-v1.0-8-bit quantized): A Persian language model prompted directly in Persian, with outputs compared to the same gold-standard Persian counter-narratives.
\end{itemize}

\textbf{Evaluation Metrics:}
\begin{itemize}
\item \textbf{Automated Metrics:}
\begin{itemize}
\item \textbf{Standard Metrics}: BLEU, ROUGE-L, METEOR (measuring lexical and structural similarity), and BERTScore F1 (measuring semantic similarity).
\item \textbf{Quality and Safety Metrics}: Perplexity (fluency, lower is better; measured using a GPT-2 Persian model for PersianMind and a standard language model for mBART) and toxicity (lower is safer, measured with \texttt{texttox/glot500-toxicity-classifier}~\cite{imani2023glot500} for PersianMind).
\end{itemize}
\item \textbf{Human Evaluation Metrics} (for mBART and PersianMind, scored on a 1--5 scale, higher is better):
\begin{itemize}
\item \textbf{Relevance}: How well the response addresses the hate speech content.
\item \textbf{Effectiveness}: The response’s ability to mitigate hate speech or promote constructive dialogue.
\item \textbf{Tone Appropriateness}: Suitability of the tone for counter-narrative responses (e.g., empathetic, non-confrontational).
\item \textbf{Fluency}: Linguistic coherence and naturalness of the response.
\item \textbf{Cultural Relevance}: Alignment with Persian cultural norms and context.
\end{itemize}
\end{itemize}

\subsection{Results}
The performance of the evaluated models—mBART and PersianMind—is summarized in Tables \ref{tab:standard_metrics}, \ref{tab:quality_safety}, and \ref{tab:human_metrics}.

Lexical and Structural Similarity:
As shown in Table \ref{tab:standard_metrics}, both models exhibit extremely low BLEU, ROUGE-L, and METEOR scores, highlighting minimal overlap with the gold-standard references in terms of lexical choices or syntactic structure. mBART performs slightly better (BLEU: 0.0096, ROUGE-L: 0.0855, METEOR: 0.0892), but these values remain insufficient for practical use. PersianMind fails to achieve any non-zero score on BLEU and ROUGE-L, with a low METEOR score (0.0626), indicating limited synonym or stem-level alignment.

Semantic Similarity:
Despite weak lexical resemblance, both models achieve relatively high BERTScore F1 values (mBART: 0.8464, PersianMind: 0.8305), suggesting some semantic proximity to the gold references. However, this similarity is primarily due to the models generating generic counter-narrative with broadly positive sentiments (e.g., promoting empathy and peace). These outputs often lack the contextual specificity and cultural grounding necessary for persuasive, situation-sensitive counter-narrative.

Fluency and Safety (Automated Metrics):
As shown in Table \ref{tab:quality_safety}, mBART demonstrates the lowest perplexity (65.73), indicating high fluency, though this may come at the cost of generating templated or repetitive text. PersianMind, while more fluent than other unreported baselines, still exhibits a high perplexity score (316.35), far above the gold Persian references (116.17), suggesting limited coherence and linguistic naturalness.
On the safety front, mBART’s outputs are notably more toxic (0.2640) than human-authored gold responses (0.0163 in English, 0.0598 in Persian), raising red flags for deployment in sensitive domains. PersianMind performs better (toxicity: 0.1157), but still exceeds safe thresholds compared to human benchmarks, signaling a risk of inappropriate or counterproductive content generation.

Human Evaluation:
Table \ref{tab:human_metrics} presents a detailed comparison of human ratings across five dimensions.

Relevance and Effectiveness: Both models received low scores, with mBART scoring 2.35 for relevance and 1.86 for effectiveness, and PersianMind at 2.24 and 2.02, respectively. These results reflect the models’ inability to directly and constructively address hate speech content.

Tone Appropriateness: PersianMind outperformed mBART (3.41 vs. 2.49), showing a relatively more empathetic and non-confrontational tone, yet still lagging behind the gold standard (4.34).

Fluency: Both models demonstrated moderate fluency (mBART: 3.30, PersianMind: 3.58), aligning with automated perplexity trends. However, both remained below human-authored responses (4.61).

Cultural Relevance: Scores for cultural relevance were low across the board (mBART: 1.95, PersianMind: 2.08), underscoring a disconnect between generated outputs and Persian socio-cultural norms. This gap limits the models’ utility in real-world applications, where cultural sensitivity is paramount.

Overall Assessment:
These results collectively highlight serious limitations in current baseline systems for Persian counter-narrative generation. The outputs are often semantically generic, lexically distant from human responses, and culturally misaligned. Moreover, high toxicity levels and poor human evaluation scores raise concerns about both safety and efficacy.
These deficiencies primarily stem from the lack of Persian-specific training data. While mBART relies on English training data and cross-lingual transfer, it fails to capture Persian nuance even with high-quality translations. PersianMind, despite being monolingual, lacks fine-tuning on domain-specific counter-narrative tasks, resulting in bland or misdirected outputs.

The findings emphasize the need for culturally anchored, Persian-language resources like ParsCN to train models that can generate precise, fluent, safe, and context-aware counter-narratives for online hate mitigation.

\begin{table}[h]
\centering
\small
\caption{Standard Evaluation Metrics for Model Generations vs. Gold References}
\label{tab:standard_metrics}
\begin{tabular}{lcc}
\toprule
\textbf{Metric} & \textbf{mBART} & \textbf{PersianMind} \\
\midrule
BLEU & 0.0096 & 0.0000 \\
ROUGE-L & 0.0855 & 0.0000 \\
METEOR & 0.0892 & 0.0626 \\
BERTScore F1 & 0.8464 & 0.8305 \\
\bottomrule
\end{tabular}
\end{table}

\begin{table}[h]
\centering
\footnotesize
\caption{Quality and Safety Metrics for Model Generations vs. Gold References}
\label{tab:quality_safety}
\resizebox{\columnwidth}{!}{
\begin{tabular}{lccc}
\toprule
\textbf{Metric} & \textbf{mBART} & \textbf{PersianMind} & \textbf{Gold References} \\
\midrule
Perplexity & \textbf{65.73} & 316.35 & \textbf{102.23 (En), 116.17 (Fa)} \\
Toxicity & 0.2640 & 0.1157 & \textbf{0.0163 (En), 0.0598 (Fa)} \\
\bottomrule
\end{tabular}}
\end{table}

\begin{table}[h]
\centering
\footnotesize
\caption{Human Evaluation Metrics for Model Generations vs. Gold References (1--5 Scale)}
\label{tab:human_metrics}
\resizebox{\columnwidth}{!}{
\begin{tabular}{lccc}
\toprule
\textbf{Metric} & \textbf{mBART} & \textbf{PersianMind} & \textbf{Gold References} \\
\midrule
Relevance & 2.35 & 2.24 & \textbf{3.97} \\
Effectiveness & 1.86 & 2.02 & \textbf{4.04} \\
Tone Appropriateness & 1.49 & 3.41 & \textbf{4.34} \\
Fluency & 3.30 & 3.58 & \textbf{4.61} \\
Cultural Relevance & 1.95 & 2.08 & \textbf{3.68} \\
\bottomrule
\end{tabular}}
\end{table}

\section{Appendix E: Examples of Culturally Relevant Samples}

This appendix provides examples of hate speech and counter-narrative pairs from MultiTarget-CONAN and HateXplain, selected and translated to align with Persian socio-cultural contexts, as described in the Data Sourcing section. These examples, presented in both Persian and English, demonstrate the cultural relevance of the selected samples, addressing concerns about their fit within Persian online discourse. Figure \ref{fig:translated-examples} presents three examples from each dataset, with hate speech and corresponding counter-narratives in Persian and their English translations.

\subsection{Relevance to Persian Context}

The examples in Figure \ref{fig:translated-examples} were selected and post-edited to align with socio-cultural issues prevalent in Persian online discourse, ensuring their relevance to Persian-speaking communities. Below, we explain how each example reflects specific cultural and social dynamics in Persian contexts:

\begin{itemize}
    \item \textbf{MultiTarget-CONAN Examples (Gender)}: The three examples targeting women address gender-based discrimination, a prominent issue in Persian online spaces. The hate speech samples reflect common misogynistic narratives in Persian discourse, such as dismissing women’s political participation or confining them to domestic roles, which resonate with ongoing debates about gender equality in Iran and other Persian-speaking regions. For instance, the claim that “women have no place in politics” mirrors sentiments often expressed in Persian social media, where traditional gender roles are debated. The counter-narratives, emphasizing equality and women’s capabilities, align with progressive movements advocating for women’s rights in public and political spheres, making them highly relevant to Persian audiences.
    \item \textbf{HateXplain Examples (National)}: The three examples targeting Arabs and Chinese individuals reflect ethnic and geopolitical tensions prevalent in Persian online discourse. Anti-Arab sentiment, as seen in references to “fewer Arabs” or “filthy Arab country,” is rooted in historical and political rivalries in the Middle East, often amplified in Persian social media due to regional conflicts (e.g., Iran’s relations with Arab states). Similarly, the derogatory “ching chong” remark targeting Chinese individuals reflects stereotypes that emerge in Persian online spaces, influenced by global media and geopolitical perceptions of China. The counter-narratives, which promote humanity, peace, and character-based judgment, resonate with Persian cultural values of hospitality and respect, countering hate with calls for unity that are contextually appropriate for Persian-speaking communities.
\end{itemize}

These examples were carefully selected to address issues like gender discrimination and ethnic prejudice, which are prevalent in Persian online discourse. Native Persian annotators post-edited the translated samples to ensure linguistic accuracy and cultural congruence, as evidenced by high human evaluation scores for relevance (4.025--4.231) and tone appropriateness (4.507--4.787) (Table \ref{tab:human_evaluation_scores}). This process ensures that the ParsCN dataset captures the ``Persian soul'' by reflecting authentic socio-cultural dynamics.

\section{Appendix F: Annotator Details}
All annotators are native Persian speakers with graduate or doctoral backgrounds in Computational Linguistics or Social Computing, each with prior experience in hate speech or counter-narrative research. Their academic and linguistic backgrounds ensured a deep familiarity with the socio-cultural dynamics relevant to Persian online discourse. Annotators were based in Persian-speaking regions (including major urban centers like Tehran and Qazvin), with a balanced gender distribution and an age range between early 20s and early 30s. Before annotation, they completed a two-hour calibration workshop and a qualification quiz (20 samples) to ensure conceptual alignment. Inter-annotator agreement reached Cohen’s $\kappa = 0.71$ (substantial), and semantic consistency on shared samples achieved a BERTScore $F1 = 0.82$, confirming both reliability and coherence. Ethical approval was obtained under institutional research guidelines, with full anonymization and voluntary participation.

\section{Appendix G: Ethical Considerations}
The development of resources to combat hate speech must carefully address ethical considerations. Given that the dataset contains sensitive and potentially offensive content, strict protocols for storage, access, and distribution are essential to prevent misuse and ensure researcher safety. Bias is a significant concern, stemming from the source data, annotation decisions, and LLMs utilized; this would result in datasets or models that unintentionally under-represent specific types of hate speech or generate biased or ineffective counter-narratives to specific target groups. Finding the right balance between effective counter-narrative and avoiding censorship is a particularly sensitive challenge; it's vital that systems trained on resources like ParsCN are designed to promote healthy dialogue without suppressing legitimate, critical expression. Finally, the actual effectiveness and real impact of automatically generated counter-narratives in countering online hate speech is a complex problem that calls for further investigation. The question needs to go beyond simply generating syntactically correct answers to analyzing their social and practical implications in the specific Persian online culture, and also considering the responsibility for the generated outputs.

\section{Appendix H: Qualitative Error Analysis}
\label{app:error_analysis}
A qualitative inspection of the automatically generated counter-narratives revealed several recurring failure modes prior to manual curation. First, some responses were overly generic and failed to engage with the specific hateful claim, instead producing broad appeals to kindness or peace. Second, some fact-based responses introduced unsupported or weakly grounded claims, which risked sounding hallucinatory or unpersuasive. Third, a subset of outputs exhibited cultural misalignment, including phrasing that was grammatically acceptable in Persian but pragmatically unnatural or insufficiently adapted to Persian socio-political discourse. Fourth, some responses used an overly confrontational or sarcastic tone, which conflicted with our goal of promoting constructive and non-escalatory counter-narrative. Manual curation addressed these issues by revising factual content, improving specificity, adapting culturally incongruent wording, and softening tone where necessary. This analysis further supports the importance of human oversight in low-resource, culturally sensitive counter-narrative generation.

\subsection{Appendix I: Comparison of Native Persian and Translated-Source Pairs}
To further examine whether translated source material introduced noticeable degradation in the final Persian hate speech--counter-narrative pairs, we conducted a focused comparative human evaluation between native Persian pairs derived from PHATE and translated-and-adapted pairs derived from MT-CONAN. Since MT-CONAN was incorporated only for the overlapping Gender and Racial target groups in Stage One, we restrict this analysis to those categories to ensure a controlled comparison between native and translated-source content.

We randomly sampled a balanced set of pairs from the two sources and asked two native Persian annotators to independently evaluate each hate speech--counter-narrative pair as a whole. Rather than evaluating the hate speech and counter-narrative separately, annotators scored the pair-level quality using four criteria: \textit{Cultural Naturalness}, \textit{Semantic Clarity}, \textit{Fluency}, and \textit{Tone Appropriateness}. Cultural Naturalness measures whether the pair feels plausible and natural in contemporary Persian discourse; Semantic Clarity assesses whether the meaning of the hate speech, the response, and their relationship are clear and coherent; Fluency evaluates the overall linguistic well-formedness of the pair; and Tone Appropriateness captures whether the counter-narrative responds in a respectful and contextually suitable manner. All scores were assigned on a 1--5 scale.

Table~\ref{tab:phate_mtconan_pair_comparison} reports the average scores for each annotator and the combined mean. The PHATE pairs achieve slightly higher scores in Cultural Naturalness (4.14 vs.\ 3.75) and Tone Appropriateness (4.43 vs.\ 4.18), while translated-source pairs remain highly competitive in Fluency (4.26 vs.\ 4.30) and slightly exceed PHATE in Semantic Clarity (4.07 vs.\ 3.98). Overall, the translated-and-adapted pairs receive strong scores across all four dimensions, indicating that the translation and manual post-editing pipeline preserved high quality in the final Persian dataset.

These results support our design choice of using translation followed by careful human adaptation for low-resource dataset construction. Although native Persian pairs remain somewhat more culturally natural on average, the gap is modest, and the translated-source pairs remain fluent, semantically clear, and tone-appropriate. This suggests that our localization process substantially mitigated translation artifacts and enabled translated examples to serve as reliable additions to the dataset, especially in categories where native Persian coverage was limited. Importantly, the translated-source pairs do not exhibit a broad degradation in quality: their scores remain above 4.0 in Semantic Clarity, Fluency, and Tone Appropriateness, and only Cultural Naturalness shows a noticeable but limited decrease relative to native Persian pairs. We therefore view translation followed by culturally informed human adaptation as a practical and effective strategy for expanding Persian counter-narrative resources when native data is scarce.

\begin{table}[t]
\centering
\small
\caption{Pair-level human evaluation comparing native Persian PHATE pairs and translated-and-adapted MT-CONAN pairs. Scores are on a 1--5 scale.}
\label{tab:phate_mtconan_pair_comparison}
\resizebox{.98\columnwidth}{!}{
\begin{tabular}{lccccccccccccc}
\toprule
\multirow{2}{*}{\textbf{Source}} 
& \multicolumn{3}{c}{\textbf{Cultural Naturalness}} 
& \multicolumn{3}{c}{\textbf{Semantic Clarity}} 
& \multicolumn{3}{c}{\textbf{Fluency}} 
& \multicolumn{3}{c}{\textbf{Tone Appropriateness}} \\
\cmidrule(lr){2-4}\cmidrule(lr){5-7}\cmidrule(lr){8-10}\cmidrule(lr){11-13}
& A1 & A2 & Avg & A1 & A2 & Avg & A1 & A2 & Avg & A1 & A2 & Avg \\
\midrule
PHATE      & 4.250 & 4.033 & 4.142 & 4.100 & 3.867 & 3.983 & 4.417 & 4.183 & 4.300 & 4.517 & 4.333 & 4.425 \\
Translated & 3.767 & 3.733 & 3.750 & 4.150 & 3.983 & 4.067 & 4.350 & 4.167 & 4.258 & 4.250 & 4.100 & 4.175 \\
\bottomrule
\end{tabular}}
\end{table}

\section{Appendix J: Limitations}
\label{sec:limitations}

While introducing the first Persian counter-narrative dataset, this study faces some limitations. The 1100 pair dataset size, while substantial for a low-resource language baseline, is smaller than for high-resource languages and may possibly limit model generalizability. Relying on translated and adapted hate speech examples from English datasets, despite manual post-editing, cannot fully capture the entire scope and varieties of native Persian online hate speech and counter-narratives. Furthermore, the utilization of LLMs for generation in Stage Two, for all its scalability, means that generated counter-narratives can reflect biases present in the LLM training data or miss out on the spontaneity and contextual richness of actual human responses, even following manual curation. In addition, our evaluation is limited to offline human and automatic assessment; we do not study how generated counter-narratives perform in live online environments or whether they influence user beliefs, engagement, or de-escalation outcomes.

\section{Appendix K: AI assistance in Writing}
AI Assistants were used solely to assist in improving the writing clarity and language of this paper. Specifically, AI-assisted refinements were applied to enhance readability, coherence, and grammatical accuracy. No AI-generated content was used to replace critical thinking or fabricate results. Ideas, methodology, experimental design, analysis, and conclusions were entirely conceived, developed, and executed by the authors.

\end{document}